\documentclass[prl,twocolumn, superscriptaddress, showpacs]{revtex4}

\usepackage{graphicx} \usepackage{amssymb} \usepackage[usenames]{color}

\begin{document}





%
%

\title{Strain control of superlattice implies weak charge-lattice coupling in
La$_{0.5}$Ca$_{0.5}$MnO$_3$}

\author{S. Cox} \affiliation{Department of Materials Science and Metallurgy,
University of Cambridge, Cambridge, CB2 3QZ, UK} \author{E. Rosten}
\affiliation{Department of Engineering, University of Cambridge, Cambridge CB2
1PZ, UK} \author{J.C. Chapman} \affiliation{Department of Materials Science and
Metallurgy, University of Cambridge, Cambridge, CB2 3QZ, UK} \author{S.
Kos}\affiliation{Cavendish Laboratory, Cambridge, CB3 0HE, UK} \author{M.J.
Calder\'{o}n} \affiliation{Cavendish Laboratory, Cambridge, CB3 0HE, UK}
\affiliation{Condensed Matter Theory Center, Department of Physics, University
of Maryland, College Park, Maryland 20742-4111} \author{D.-J.~Kang}
\affiliation{Nanoscience Centre, University of Cambridge, Cambridge, CB3 0FF,
UK} \affiliation{Sungkyunkwan Advanced Institute of Nanotechnology,
Sungkyunkwan University, Suwon 440-746, Korea} \author{P.B.  Littlewood}
\affiliation{Cavendish Laboratory, Cambridge, CB3 0HE, UK} \author{P.A.
Midgley} \author{N.D.  Mathur}\email{ndm12@cus.cam.ac.uk}
\affiliation{Department of Materials Science and Metallurgy, University of
Cambridge, Cambridge, CB2 3QZ, UK}

\begin{abstract}

We have recently argued that manganites do not possess stripes of charge order,
implying that the electron-lattice coupling is weak [Phys Rev Lett \textbf{94}
(2005) 097202].  Here we independently argue the same conclusion based on
transmission electron microscopy measurements of a nanopatterned epitaxial film
of La$_{0.5}$Ca$_{0.5}$MnO$_3$.  In strain relaxed regions, the superlattice
period is modified by 2-3\% with respect to the parent lattice, suggesting that
the two are not strongly tied.

\end{abstract}


\maketitle

The superlattice present in many manganites has traditionally been described in
terms of a charge ordered array of the idealised cations Mn$^{3+}$ and
Mn$^{4+}$~\cite{Goodenough, Wollan, chen_comm_incomm,neil_ssc}.  This
superlattice is observed in x-ray, neutron and electron diffraction patterns as
extra reflections that typically lie along or near \textbf{a}$^*$, indexing the
room temperature cell as orthorhombic $Pnma$.  Recent work controversially
suggests that Mn valence charges are not strongly localised, and that any
charge modulation is very small~\cite{garcia,rodcar,valeria,us,brey,proffen}.

We recently argued that in polycrystalline La$_{1-x}$Ca$_x$MnO$_3$ ($x>0.5$) at
90 K, the charge-lattice coupling is weak because the superlattice is not
locked to the parent lattice~\cite{us}.  Instead, the periodicity of the
superlattice was found to be uniform over a wide range of length scales in any
particular grain. Our main evidence was that the superlattice wavenumber $q$
was invariant with respect to $a^*$ when a grain was repeatedly sampled with a
local probe (convergent beam electron diffraction, spot size 3.6 nm).  This
interpretation relied upon selecting x=0.52 such that $q/a^* \approx
1-x$~\cite{chen_incomm} was near but not equal to 0.5.  In bulk unstrained
La$_{0.5}$Ca$_{0.5}$MnO$_3$, $q/a^*=0.5$ below the N\'{e}el transition
temperature $T_N\sim135$ K (on cooling)~\cite{chen_comm_incomm}. The
superstructure persists up to the Curie temperature of $T_C\sim220$ K, and for
$T_N<T<T_C$, $q/a^*$ is hysteretic and incommensurate~\cite{chen_comm_incomm}.

It has previously been suggested that the superlattice of a manganite should be
modified by strain~\cite{neil_ssc}.  Intergranular variations in $q/a^*$ of up
to 8\% have been observed in polycrystalline
La$_{0.5}$Ca$_{0.5}$MnO$_3$~\cite{philmag}, but the possibility of extrinsic
effects precludes a direct link with strain.  Here we investigate tuning the
strain state in a continuous crystal lattice, where extrinsic effects should be
minimised.  Although chemical phase separation prevents the growth of bulk
single crystal La$_{1-x}$Ca$_x$MnO$_3$ ($x\geq0.41)$~\cite{crystal},  we have
formed an untwinned continuous crystal lattice by growing a coherently strained
epitaxial film of  La$_{0.5}$Ca$_{0.5}$MnO$_3$ on an orthorhombic
NdGaO$_3$~(001) substrate (NGO).  Superlattice reflections are expected to be
strongest at this composition, since optical spectroscopy measurements show a
``pseudogap'' in La$_{1-x}$Ca$_{x}$MnO$_3$ that is largest at
$x=0.5$~\cite{pseudogap}.  We have attempted to release the epitaxial strain in
some areas of the film by firstly removing substrate material to create an
electron transparent window $\sim$150 nm thick, and then removing material
around rectangular micron-scale regions (``rectangles'') within the window.
Transmission electron microscopy (TEM) revealed that $q/a^*$ is reduced by
2-3\% inside the rectangle.

Films were grown at $\sim$800$^\circ$C in a flowing oxygen ambient of 15~Pa by
pulsed laser deposition from a polycrystalline La$_{0.5}$Ca$_{0.5}$MnO$_3$
target (Praxair, USA) using a 248~nm ultraviolet KrF laser with an average
fluence of 1.5~J.cm$^{-2}$, a repetition rate of 1~Hz and a target-substrate
distance of 8~cm.  Films were subsequently annealed for one hour in 60 kPa
O$_2$ at $\sim$800$^o$C.  The $a$ lattice parameter of NGO at the 90 K nominal
base temperature of our microscope stage is 0.48\% smaller than the $a$ lattice
parameter of La$_{0.5}$Ca$_{0.5}$MnO$_3$, and the mismatch in $b$ is 0.35\% in
the opposite sense~\cite{ngo_temp}.  The film was $44\pm2$~nm thick as measured
by high resolution X-ray diffraction (HRXRD).  This thickness is sufficiently
low to preserve cube-on-cube epitaxy.  An X-ray rocking curve with a FWHM of
0.10$^\circ$ for the (004) film reflection was recorded, and a typical value
for surface roughness as measured by atomic force microscopy was $\sim$0.5 nm.
A ferromagnetic signal detected below room temperature reached an apparent
saturation magnetization of 0.6~$\mu_B$/Mn at 90 K, with no evidence for the
antiferromagnetic transition that is observed in the bulk above 100
K~\cite{Schiffer}. Similarly, no transitions were seen in the electrical
resistivity, which was 0.02~$\Omega$.cm at 300 K and remained insulating down
to 80 K, beyond which we could no longer measure it.

\begin{figure} \begin{centering}
\includegraphics[width=0.45\textwidth]{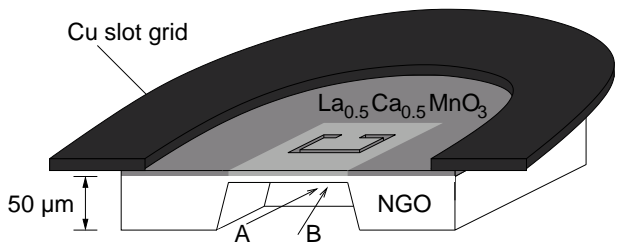}
\caption{Sample preparation of a ``rectangle'' in an FIB microscope.  A beam of
Ga ions in direction ``A'' was used to mill away 22 $\mu$m $\times$ 8 $\mu$m of
substrate from underneath the film.  A beam of Ga ions in direction ``B'' was
then used to mill cuts, delineated with thin black lines in the light grey
region.  This light grey region represents the $\sim$150 nm thick electron
transparent window.  The dark grey region represents film underneath which
50~$\mu$m of substrate remains.  The sample was attatched with silver glue to
half of a TEM Cu grid support with an outer diameter of 3 $\mathrm{mm}$.
\label{FIB}}
\end{centering} \end{figure}

\begin{figure} \begin{centering}
\includegraphics[width=0.5\textwidth]{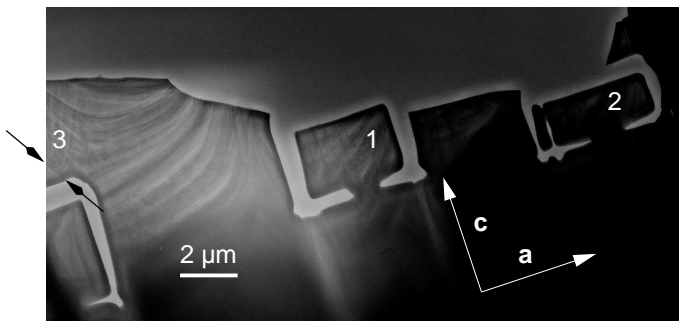} \caption{TEM image of
Rectangle 1 and Rectangle 2.  The material directly above the rectangles has
broken away.  A crack runs between and parallel to the arrows in region
3.\label{rectangles}} \end{centering} \end{figure}

The sample was prepared for TEM by conventional grinding to 50 $\mu$m, and
processing using the focussed ion beam (FIB) microscope (Fig.~\ref{FIB}). The
electron transparent window was defined by cutting substrate material from
under the film.  When the window was $\sim1$ $\mu$m thick, the  sample was
tilted 45$^\circ$ and cuts were made from the substrate side to minimise film
damage.  These cuts defined a free standing rectangular region (a
``rectangle'').  The sample was then rotated back to its original position with
sufficient precision to avoid an undercut during subsequent thinning of the
window to electron transparency. Material furthest from the front edge of the
window in Fig.~\ref{FIB} was therefore thickest.  A low magnification TEM
picture of two rectangles is shown in Fig.~\ref{rectangles}.  The minimum
thickness of the window that could be achieved reliably was $\sim$150~nm.  Thus
$\sim$100~nm of substrate remained attached to the 44 nm film.

The sample was cooled to approximately 90 K for up to four hours at a time
using a Gatan double-tilt liquid nitrogen stage.  Parent lattice reflections
were recorded in diffraction patterns with a CCD camera on a Philips CM300 TEM
operated at 300~kV.  However, superlattice reflections were too weak to measure
on the CCD without significant over-saturation of the parent reflections.
Therefore measurements of $q/a^*$ were extracted from diffraction patterns
recorded on photographic film, which has a sensitive nonlinear response.  For
this a Philips CM30 TEM operated at 300~kV was used with a 500~nm aperture.

At 90 K all regions of the electron transparent window (both inside and outside
the rectangles) produced diffraction patterns showing the superlattice.  As
expected, the superlattice modulations were always parallel or near-parallel to
the \textbf{a}$^*$ direction.  Custom written software was used in order to
measure statistically significant values of $q/a^*$ for each diffraction
pattern.  Initially the parent lattice reflections were identified and the
distortion of the photographic film was calculated, then the positions of the
superlattice reflections were found.  Thus values of $q/a^*$ were established
for each diffraction pattern.

Specifically, the positions of the parent lattice reflections  were estimated
and then refined using the mean-shift algorithm.  The film distortion was
calculated using the projective warp which models the distortion as shear,
aspect ratio change and keystoning.

Pairs of superlattice reflections that appeared between adjacent pairs of
parent lattice reflections along the \textbf{a}$^*$ axis were modelled using
the weighted sum of two Gaussians and a constant value.  The parameters were
fitted to this Gaussian Mixture Model (GMM) using the Expectation Maximisation
algorithm~\cite{em, gaussian}.  Information was ignored from areas near the
edge of the photographic film that were warped such that the mismatch between
the expected lattice and the observed lattice was greater than two pixels.  The
curvature of the Ewald sphere leads to a systematic error $\sim(g/k)^2$, where
$g$ is the measured value of the wavevector and $k$ is the wavevector measured
across the Ewald sphere, but this is small and will affect equally both the
parent and superlattice reflections, such that it may be ignored here.

\begin{figure} \begin{centering}
\includegraphics[width=0.45\textwidth]{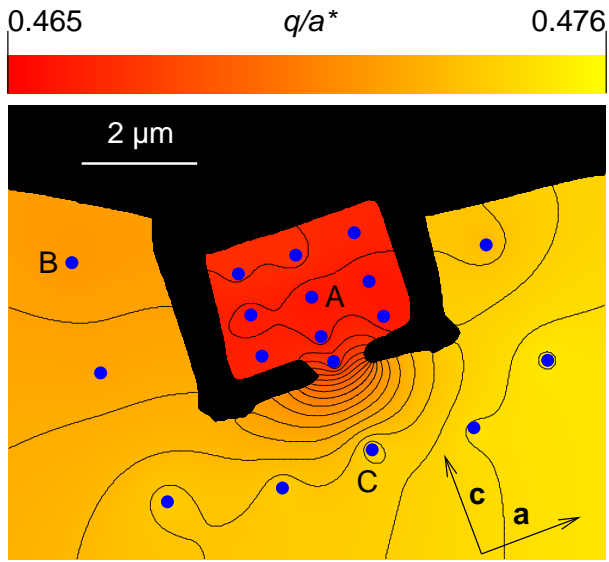}
\caption{(color online) False colour map of $q/a^*$ at 90 K in and around
Rectangle 1 with contours of constant $q/a^*$ plotted every
$\Delta(q/a^*)=5.8\times10^{-4}$.  Diffraction patterns were taken at the 18
points indicated, and $q/a^*$ values were extracted from each using the
software described in the text.  Data for $q/a^*$ was generated away from the
18 points by interpolation and extrapolation.  The diagram combines data from
four cooling runs to 90 K.  One run included data from A, B and C and other
runs included data from at least one of these points.  Data from the other runs
was subject to the run to run variations described in the text.  It was
therefore offset to build the above picture.  Thermal drift is estimated to be
0.2 nm.\label{colourmap}} \end{centering} \end{figure}

Fig.~\ref{colourmap} shows a map of $q/a^*$ in and around Rectangle 1.  The
magnitude of $q/a^*$ was highest at C, 0.8\% lower at B, and 1.4\% lower again
inside the rectangle at A (0.4760$\pm$0.0009, 0.4710$\pm$0.0005 and
0.4646$\pm$0.0006, respectively). Similarly, for Rectangle~2, $q/a^*$ at points
analogous to B and A differ in the same sense by 1.3\% (0.4753$\pm$0.0005 and
0.4692$\pm$0.0007 respectively).  In any given diffraction pattern, each
individual measurement of $q/a^*$ was recorded to within 0.004, given a
resolution of 0.3 out of 35 pixels.  For each diffraction pattern, between 150
and 300 measurements of $q/a^*$ were made, reducing this error to the values
quoted.

At any point in the window, the measured wavenumber varied between cooling
runs.  The range of $q/a^*$ inside Rectangle 1 at point A was 2.6\% (0.457 --
0.469).  Outside Rectangle 1 at point C, the range was 1.9\%  (0.467 -- 0.476).
However, in any given run, the wavenumber outside the rectangle was always
larger than the wavenumber inside the rectangle, with the run to run difference
from A to C being between 2.2\% -- 3.2\%.

Since $q/a^*$ rather than $q$ is measured, we investigated whether the observed
variations of a few $\%$ could be due to variations in $a^*$ alone.  The parent
lattice reflections were recorded in different areas of the sample above and
below the ordering transition temperature of $\sim$220~K as estimated from
polycrystalline samples~\cite{chen_comm_incomm}.  Variations in $a^*/c^*$ were
$\leq1\%$, which assuming $c$ to be constant implies that variations in $a^*
\leq 1\%$.  This places an upper bound of 0.1\% on changes in $q/a^*$ due to
unresolved changes in $a^*$.  (Note that this error calculation is non-trivial
because the measured $q$ is always determined relative to the measured $a^*$.)
Therefore the spatial variations seen in $q/a^*$ represent changes in $q$,
whether or not they are driven by changes in $a^*$ that are beyond the 1\%
resolution of the microscope.

The asymmetry in $q/a^*$ with respect to the artificial cuts rules out the
possibility that contamination and/or damage from the Ga beam of the FIB
microscope produce the observed changes in our measurements taken at points
over 500nm from the artificial edges.  Moreover, when moving from 4 $\mu$m to
within 1 $\mu$m of a natural crack (Fig.~\ref{rectangles}), $q/a^*$ was reduced
by 1.3\% (0.476 to 0.470).  This mimics the change in $q/a^*$ that we
engineered in the rectangle.

The observed differences between $q/a^*$ inside and outside the rectangle could
be due to the electron beam heating the rectangle, which is thermally isolated
by its small neck.  However, one would then expect $q/a^*$ to vary in a
systematic way with remoteness from the neck. This is not the case so thermal
effects cannot explain the results of this experiment.

\begin{figure} \begin{centering}
\includegraphics[width=0.5\textwidth]{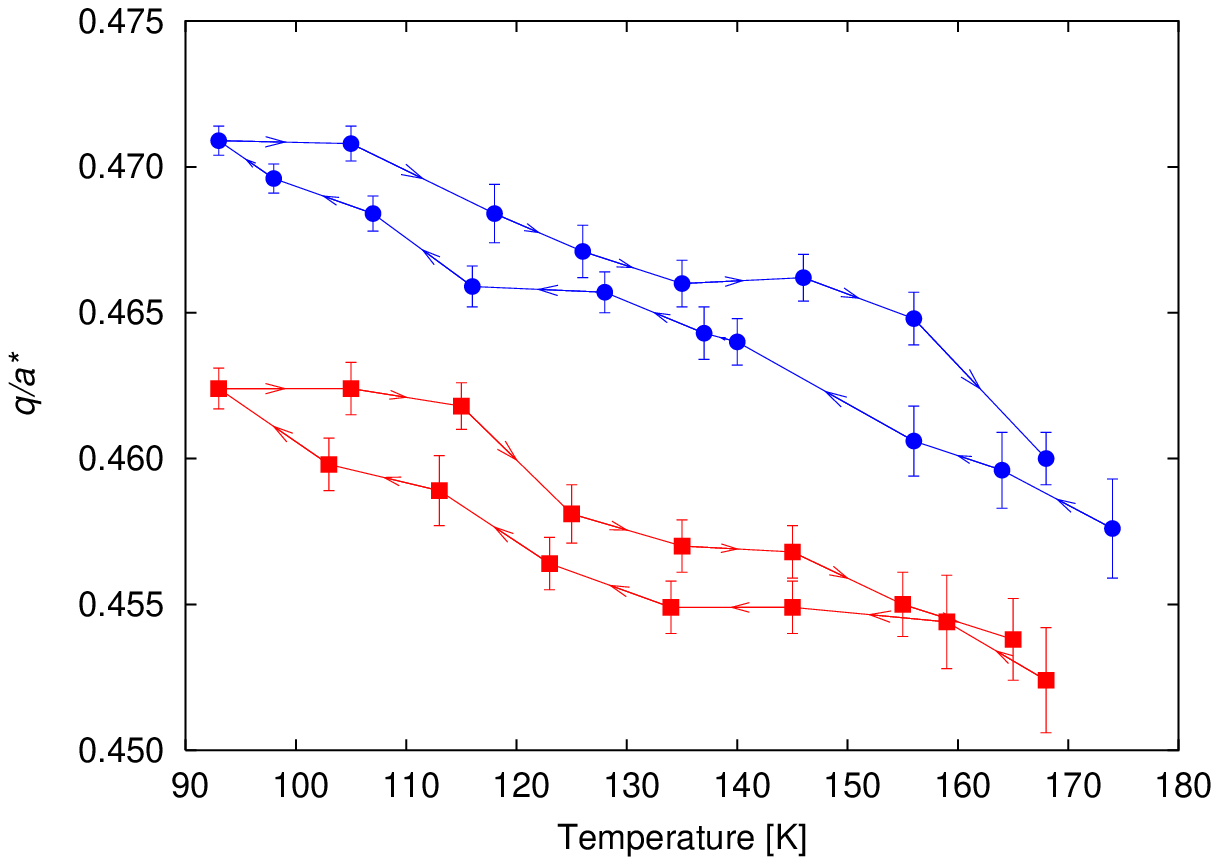}
\caption{(color online) Variation of $q/a^*$ with temperature, inside
({\tiny$\textcolor{red}{\blacksquare}$}) and outside
($\textcolor{blue}{\bullet}$) Rectangle 1. The readings were taken at A and B
using a 2 $\mu$m aperture. There is a 1 $\mu$m spatial uncertainty due to
thermal drift of the sample during data acquisition.  The error bars are at one
standard deviation of the mean.  Note that recent measurements using a Gatan
helium stage suggest that the two 90 K values remain constant within error down
to $\sim$15 K.
\label{hyst}} \end{centering} \end{figure}

The observed reduction of $q/a^*$ inside the rectangle could also arise if
discommensurations, which separate regions of different $q/a^*$, were pinned
strongly inside the rectangle, due to defects at the nearby edges, and could
not propagate through the neck.  Temperature sweeps taken inside and outside
the rectangle both show a similar hysteresis of $\sim$20 K (Fig.~\ref{hyst}).
This suggests that the degree of pinning is similar inside and outside the
rectangle, and that pinning does not cause the observed differences in $q/a^*$.

We suggest that small changes in strain, below our $1\%$ resolution in $a^*$,
are responsible for the observed variations in $q/a^*$.  Indeed, changes this
small can be significant.  For example, a 0.5\% change of strain~\cite{Soh2}
along the normal to the surface of a La$_{0.7}$Sr$_{0.3}$MnO$_3$~\cite{Soh}
film produces a 20 K change in the Curie temperature.

Our observation that $q/a^*$ is smallest inside the rectangle may be understood
using a 1D Ginzburg-Landau theory~\cite{Landau}.  In the modulated manganite we
studied, the nature of the order parameter $\psi(\textbf r)$ is not
established~\cite{us}.  Here we express it in terms of the corresponding order
parameter $\psi_0 (\textbf{r})$ in the absence of modulations as
$\psi(\textbf{r}) = \psi_0(\textbf{r})e^{i(\textbf{Q}_c.\textbf{r} +
\phi(\textbf{r}))}$ where \textbf{r} is the spatial coordinate, $\textbf{Q}_c$
is a vector commensurate with the lattice and $\phi$ incorporates
incommensurability~\cite{milward}. The wavevector is given by $\textbf{q} =
\textbf{Q}_c + \langle \nabla \phi \rangle$, where $\langle\nabla \phi\rangle$
is the deviation of the wavevector from the commensurate value.  Therefore in
our material $q = 0.5a^* + \langle \nabla\phi \rangle$. Assuming that
$\psi_0(\textbf{r})$ is constant, we can write the free energy density for the
modulation and its coupling with strain $\eta$ as~\cite{Landau}
\begin{equation}
{\cal F}= {{\xi^2}\over{2}} \left(\nabla\phi- \delta  \right)^2+{{v}\over{n}} cos(n
\phi)+ c \eta \nabla\phi +{{1}\over{2}} \kappa \eta^2 -\sigma \eta.
\label{eq:free-energy} \end {equation}

The first term is the elastic term that favours incommensurate modulation, and
we arbitrarily set $\xi$=1.  The parameter $\delta$ is the deviation of $q/a^*$
from 0.5 in the absence of strain coupling.  We always see $q/a^*<0.5$ in our
film, which we suggest is due to the presence of a background strain that
arises from our inability to completely remove strain everywhere, in effect
rendering $\delta<0$.  The second term is the Umklapp term that favours
commensurability, where $n$ is an integer and the coefficient $v$ determines
the strength of the effect.  The third term couples $\eta$ and $\nabla \phi$
with strength $c$.  The fourth term is the strain energy density in terms of
the bulk elastic modulus $\kappa$.  The fifth term gives the elastic energy due
to the stress $\sigma$ on the film from the substrate.  The effect of the
coupling term $c \eta \nabla \phi$ on the wavevector can be determined in the
plane-wave limit ($\nabla \phi=$constant and $\nabla \eta=0$) by minimising
(\ref{eq:free-energy}), which leads to
\begin{equation} \nabla \phi = {{\delta-c \sigma/\kappa}\over{1-c^2/\kappa}}.
\end{equation}

Two limiting cases represent the situation inside and outside the rectangle
respectively: either the film relaxes in the absence of substrate-induced
stress and $q$ is reduced by $|\nabla
\phi_{in}|={{|\delta|}\over{1-c^2/\kappa}}$ to give $q = 0.5a^* -
{{|\delta|}\over{1-c^2/\kappa}}$, or the film is clamped such that the coupling
$c\eta \nabla \phi$ is inactive, and thus $|\nabla \phi_{out}|=|\delta|$ and
$q=0.5a^* - |\delta|$.  Since $|\nabla \phi_{in}| > |\nabla \phi_{out}|$ we can
understand why the deviation from the commensurate value of $q/a^*=0.5$ will be
larger inside a rectangle whatever the sign of $c$.  Note that this result is
the opposite of what might be expected given that the rectangle resembles an
unstrained single crystal.

We now consider whether the changes in $q/a^*$, that we ascribe to strain,
support our recent finding that the charge-lattice coupling is weak~\cite{us}.
In the traditional strong-coupling limit, any elastic deformation of the parent
lattice should be directly transmitted to the superlattice such that $\Delta
(q/a^*)$=0. Our finding that $\Delta (q/a^*)$=2-3\% suggests that the
superlattice can deform independently of the parent lattice. Therefore the
coupling cannot be considered arbitrarily strong.  Moreover, in the traditional
strong-coupling picture, the changes in $\Delta(q/a^*)$ that we observe would
arise due to changes in the number of [100] Mn$^{4+}$ sheets, and these are not
available at a given $x$.  In theory, our finding that $\Delta(q/a^*)\neq0$
could be explained if strain is enhanced at uncharged
discommensurations~\cite{Landau}, but discommensurations are not consistent
with a strong coupling picture at x=0.5.

In summary, we have shown that it is possible to tune the magnitude of $q/a^*$
by up to 3\% in La$_{0.5}$Ca$_{0.5}$MnO$_3$ at 90 K by processing a thin film
using an FIB microscope.  This demonstrates that tuning the microstructure of
La$_{0.5}$Ca$_{0.5}$MnO$_3$ can alter the low temperature superlattice.
Consequently the variations in wavenumber seen in polycrystalline
La$_{1-x}$Ca$_x$MnO$_3$~\cite{us, philmag} may be directly attributed to
strain. Our finding that $\Delta (q/a^*)\neq0$ may be most simply explained if
the charge and lattice are weakly coupled.  The interpretation presented here
supports our earlier suggestion~\cite{us} that a charge density wave scenario
may be appropriate.

\begin{acknowledgments} {We thank M.B. Weissman and L.E. Hueso for helpful
comments. This work was funded by the UK EPSRC, The Royal Society, the Schiff
Foundation, and Churchill College, Cambridge.}
\end{acknowledgments}

\bibliographystyle{apsrev}

\begin{thebibliography}{22} \expandafter\ifx\csname
natexlab\endcsname\relax\def\natexlab#1{#1}\fi \expandafter\ifx\csname
bibnamefont\endcsname\relax \def\bibnamefont#1{#1}\fi \expandafter\ifx\csname
bibfnamefont\endcsname\relax \def\bibfnamefont#1{#1}\fi \expandafter\ifx\csname
citenamefont\endcsname\relax \def\citenamefont#1{#1}\fi \expandafter\ifx\csname
url\endcsname\relax \def\url#1{\texttt{#1}}\fi \expandafter\ifx\csname
urlprefix\endcsname\relax\def\urlprefix{URL }\fi
\providecommand{\bibinfo}[2]{#2} \providecommand{\eprint}[2][]{\url{#2}}

\bibitem[{\citenamefont{{J.P. Goodenough}}(1955)}]{Goodenough}
\bibinfo{author}{\bibnamefont{{J.P.Goodenough}}}, \bibinfo{journal}{Phys. Rev.}
\textbf{\bibinfo{volume}{100}}, \bibinfo{pages}{564} (\bibinfo{year}{1955}).

\bibitem[{\citenamefont{{E.O. Wollan, W.C. Koehler}}(1955)}]{Wollan}
\bibinfo{author}{\bibnamefont{{E.O. Wollan, W.C. Koehler}}},
\bibinfo{journal}{Phys. Rev.} \textbf{\bibinfo{volume}{100}},
\bibinfo{pages}{1} (\bibinfo{year}{1955}).

\bibitem[{\citenamefont{{C.H. Chen, S.-W. Cheong}}(1996)}]{chen_comm_incomm}
\bibinfo{author}{\bibnamefont{{C.H. Chen, S.-W. Cheong}}},
\bibinfo{journal}{Phys. Rev. Lett.} \textbf{\bibinfo{volume}{76}},
\bibinfo{pages}{4042} (\bibinfo{year}{1996}).

\bibitem[{\citenamefont{{N.D. Mathur, P.B. Littlewood}}(2001)}]{neil_ssc}
\bibinfo{author}{\bibnamefont{{N.D. Mathur, P.B. Littlewood}}},
\bibinfo{journal}{Solid State Commun.} \textbf{\bibinfo{volume}{119}},
\bibinfo{pages}{271} (\bibinfo{year}{2001}).

\bibitem[{\citenamefont{{J. Garc\'{\i}a \emph{et al.}}}(2001)}]{garcia}
\bibinfo{author}{\bibnamefont{{J. Garc\'{\i}a \emph{et al.}}}},
\bibinfo{journal}{J. Phys.-Condens. Mat.} \textbf{\bibinfo{volume}{13}},
\bibinfo{pages}{3243} (\bibinfo{year}{2001}).

\bibitem[{\citenamefont{{J. Rodr\'{\i}guez-Carvajal \emph{et
al.}}}(2002)}]{rodcar} \bibinfo{author}{\bibnamefont{{J.
Rodr\'{\i}guez-Carvajal \emph{et al.}}}}, \bibinfo{journal}{Physica B}
\textbf{\bibinfo{volume}{320}}, \bibinfo{pages}{1} (\bibinfo{year}{2002}).

\bibitem[{\citenamefont{{V. Ferrari, M.D. Towler, P.B.
Littlewood}}(2004)}]{valeria} \bibinfo{author}{\bibnamefont{{V. Ferrari, M.D.
Towler, P.B. Littlewood}}}, \bibinfo{journal}{Phys. Rev. Lett.}
\textbf{\bibinfo{volume}{91}}, \bibinfo{pages}{227202} (\bibinfo{year}{2004}).

\bibitem[{\citenamefont{{J.C. Loudon \emph{et al.}}}(2005{\natexlab{a}})}]{us}
\bibinfo{author}{\bibnamefont{{J.C. Loudon \emph{et al.}}}},
\bibinfo{journal}{Phys. Rev. Lett.} \textbf{\bibinfo{volume}{94}},
\bibinfo{pages}{097202} (\bibinfo{year}{2005}{\natexlab{a}}).

\bibitem[{\citenamefont{{L. Brey}}(2004)}]{brey}
\bibinfo{author}{\bibnamefont{{L. Brey}}}, \bibinfo{journal}{Phys. Rev. Lett.}
\textbf{\bibinfo{volume}{92}}, \bibinfo{pages}{127202} (\bibinfo{year}{2004}).

\bibitem[{\citenamefont{{E.E. Rodriguez \emph{et al.}}}(2005)}]{proffen}
\bibinfo{author}{\bibnamefont{{E.E. Rodriguez \emph{et al.}}}},
\bibinfo{journal}{Phys. Rev. B} \textbf{\bibinfo{volume}{71}},
\bibinfo{pages}{104430} (\bibinfo{year}{2005}).

\bibitem[{\citenamefont{{C.H. Chen, S. Mori, S.-W.
Cheong}}(1999)}]{chen_incomm} \bibinfo{author}{\bibnamefont{{C.H. Chen, S.
Mori, S.-W. Cheong}}}, \bibinfo{journal}{J. Phys. IV France}
\textbf{\bibinfo{volume}{9}}, \bibinfo{pages}{Pr10-307} (\bibinfo{year}{1999}).

\bibitem[{\citenamefont{{J.C. Loudon \emph{et
al.}}}(2005{\natexlab{b}})}]{philmag} \bibinfo{author}{\bibnamefont{{J.C.
Loudon \emph{et al.}}}}, \bibinfo{journal}{Phil. Mag.}
\textbf{\bibinfo{volume}{85}}, \bibinfo{pages}{999}
(\bibinfo{year}{2005}{\natexlab{b}}).

\bibitem[{\citenamefont{{P. Majewski \emph{et al.}}}(2000)}]{crystal}
\bibinfo{author}{\bibnamefont{{P. Majewski \emph{et al.}}}},
\bibinfo{journal}{J. Mater. Res.} \textbf{\bibinfo{volume}{15}},
\bibinfo{pages}{1161} (\bibinfo{year}{2000}).

\bibitem[{\citenamefont{{K.H. Kim \emph{et al.}}}(2002)}]{pseudogap}
\bibinfo{author}{\bibnamefont{{K.H. Kim \emph{et al.}}}},
\bibinfo{journal}{Phys. Rev. Lett.} \textbf{\bibinfo{volume}{88}},
\bibinfo{pages}{167204} (\bibinfo{year}{2002}).

\bibitem[{\citenamefont{{D. Savytskii \emph{et al.}}}(2003)}]{ngo_temp}
\bibinfo{author}{\bibnamefont{{D. Savytskii \emph{et al.}}}},
\bibinfo{journal}{Phys. Rev. B} \textbf{\bibinfo{volume}{68}},
\bibinfo{pages}{024101} (\bibinfo{year}{2003}).

\bibitem[{\citenamefont{{P.E. Schiffer, A.P. Ramirez, W. Bao,
S.-W.Cheong}}(1995)}]{Schiffer} \bibinfo{author}{\bibnamefont{{P.E. Schiffer,
A.P. Ramirez, W. Bao, S.-W.Cheong}}}, \bibinfo{journal}{Phys. Rev. Lett.}
\textbf{\bibinfo{volume}{75}}, \bibinfo{pages}{3336} (\bibinfo{year}{1995}).

\bibitem[{\citenamefont{{A. Dempster, N. Laird, D. Rubin}}(1977)}]{em}
\bibinfo{author}{\bibnamefont{{A. Dempster, N. Laird, D. Rubin}}},
\bibinfo{journal}{J. Roy. Stat. Soc. B} \textbf{\bibinfo{volume}{39}},
\bibinfo{pages}{1} (\bibinfo{year}{1977}).

\bibitem[{\citenamefont{{R.A. Redner, H.F. Walker}}(1984)}]{gaussian}
\bibinfo{author}{\bibnamefont{{R.A. Redner, H.F. Walker}}},
\bibinfo{journal}{{SIAM} Review} \textbf{\bibinfo{volume}{26}},
\bibinfo{pages}{195} (\bibinfo{year}{1984}).

\bibitem[{\citenamefont{{Y.-A. Soh \emph{et al.}}}(2002)}]{Soh2}
\bibinfo{author}{\bibnamefont{{Y.-A. Soh \emph{et al.}}}}, \bibinfo{journal}{J.
Appl. Phys.} \textbf{\bibinfo{volume}{91}}, \bibinfo{pages}{7742}
(\bibinfo{year}{2022}).

\bibitem[{\citenamefont{{Y.-A. Soh \emph{et al.}}}(2001)}]{Soh}
\bibinfo{author}{\bibnamefont{{Y.-A. Soh \emph{et al.}}}},
\bibinfo{journal}{Phys. Rev. B} \textbf{\bibinfo{volume}{63}},
\bibinfo{pages}{020402} (\bibinfo{year}{2001}).

\bibitem[{\citenamefont{{P. Bak, J. Timonen}}(1978)}]{Landau}
\bibinfo{author}{\bibnamefont{{P. Bak, J. Timonen}}}, \bibinfo{journal}{J.
Phys. C Sol. State Phys.} \textbf{\bibinfo{volume}{11}}, \bibinfo{pages}{4901}
(\bibinfo{year}{1978}).

\bibitem[{\citenamefont{{G.C. Milward, M.J. Calder\'{o}n, P.B.
Littlewood}}(2005)}]{milward} \bibinfo{author}{\bibnamefont{{G.C. Milward, M.J.
Calder\'{o}n, P.B.  Littlewood}}}, \bibinfo{journal}{Nature}
\textbf{\bibinfo{volume}{433}}, \bibinfo{pages}{607} (\bibinfo{year}{2005}).

\end{thebibliography}

\end{document}